\documentclass[12pt]{article}

\usepackage{psfig}
\usepackage{overcite}
\usepackage{threeparttable}
\begin{document}

\begin{center}
\LARGE{Multiplicity of Generation, Selection, and Classification
Procedures for Jammed Hard-Particle Packings}
\vspace{0.2in}
\normalsize{}

S. Torquato$^{1,2}$ and F. H. Stillinger$^{2,3}$

\noindent$^1$ Department of Chemistry,
Princeton University, Princeton, NJ 08544 \\
$^2$ Princeton Materials
Institute, Princeton University, Princeton, NJ 08544 \\ 
$^3$ Bell
Laboratories, Lucent Technologies, Murray Hill, NJ 07974

\end{center}

\begin{center}
{\bf Abstract}
\end{center}

Hard-particle packings have served as useful starting points
to study the structure of diverse systems such
as liquids, living cells,
granular media, glasses, and amorphous solids.
Howard Reiss has played a major role in helping to illuminate
our understanding of hard-particle systems, which
still offer scientists many interesting conundrums.
Jammed configurations of hard particles are of
great fundamental and practical interest. What one
precisely means by a ``jammed'' configuration
is quite subtle and considerable ambiguity remains 
in the literature on this question. We will show
that there is a multiplicity of generation, selection, and
classification procedures for jammed
configurations of identical $d$-dimensional spheres.
 We categorize common ordered lattices according
to our definitions and discuss implications for
random disk and sphere packings. We also show 
how the concept of rigidity percolation (which
has been used to understand the mechanical properties
of network glasses) can be generalized to further
characterize hard-sphere packings.

\newpage

\section{Introduction}

The problem of packing particles
into a container or vessel of some type is one of
the oldest problems known to man. Bernal \cite{Ber65} has remarked
that ``heaps (close-packed arrangements of particles)
were the first things that were ever measured in the
form of basketfuls of grain for the purpose of trading 
or the collection of taxes.'' Today scientists study
particle packings
in order to understand
the structure of living cells, liquids, granular media, glasses
and amorphous solids, to mention but a few examples.
Since the structure of such systems is primarily determined
by the repulsive interactions between the particles, 
the hard-sphere model serves as a useful idealized
starting point for such an investigation
\cite{Re59,St64b,Re86,Re92,Re96}.

Hard spheres interact with each other only when they touch,
and then 
with an infinite repulsion reflecting their impenetrable
physical volume.
Despite the simplicity of the hard-sphere potential,
hard-sphere systems offer many conundrums, several of
which we will briefly describe. The first example
concerns the existence of an entropically driven 
disorder/order phase transition in hard-sphere and
hard-disk (two-dimensional) systems. Although there
is strong numerical evidence to support the existence of
a first-order disorder/order phase transition in three dimensions,
a rigorous proof for such a transformation is not yet available.
In two dimensions,
the state of affairs is even less certain because
it is not clear (from numerical simulations) whether
the transition, if it exists, is first-order
or a continuous Kosterlitz--Thouless--Halperin--Nelson--Young (KTHNY)
transformation \cite{Ko73,Ha78,Yo79}. 

Another conundrum involves the determination
of the densest packing of identical hard spheres.
It is only recently that a putative air-tight
rigorous proof has been devised for Kepler's conjecture:
the densest possible
packing fraction $\phi$ for identical spheres in three dimensions
is $\pi /\sqrt{18} \approx 0.7405$, corresponding
to the close-packed face-centered cubic (FCC) lattice or its stacking
variants \cite{Ha98}. Although the neighborhood grocer
would have given the same solution, proving Kepler's
conjecture is another matter. The difficulty arises
because the densest local packing is inconsistent with
global packing constraints, i.e., nonoverlapping regular
tetrahedra cannot tile space.
This is not true in two dimensions, where the 
densest local packing is consistent with the
densest global packing.

Yet another example of a conundrum concerns the venerable notion of
``random close packing'' (RCP)
of hard spheres. The traditional notion of the RCP state
is that it is the maximum
density that a large, irregular arrangement of spheres can attain
and that this density is a well-defined and unique quantity.
It was recently shown that the RCP state is in fact mathematically
ill-defined and must be replaced by a new
notion called the {\it maximally random jammed} (MRJ) state,
which can be made precise \cite{To00b}. The identification of the MRJ
state rests on the development of metrics for order
(or disorder), a very challenging problem in condensed-matter
theory, and a precise definition for the term ``jammed.'' 
Torquato et al.~\cite{To00b} have suggested and computed several scalar
metrics for order in hard-sphere systems and have
introduced a precise definition of ``jammed.''
This formalism provides a means of classifying
jammed structures in terms of their degree of order
in an ``ordering'' phase diagram\footnote{Perhaps more importantly,
the ordering phase
diagram can also serve as a means for mapping the degree of order in
nonequilibrium (or history-dependent) structures as a function of their
processing conditions.}.
For example, let $\psi$ represent a scalar order
metric that varies between unity in the case of
perfect order and zero in the case of perfect disorder, and
imagine the set of all jammed structures in the $\phi$-$\psi$
plane. The MRJ state is simply 
the configuration of particles that
minimizes $\psi$ (maximizes the disorder) among
all statistically homogeneous and isotropic {\it jammed}
structures. More generally, the identification of the boundaries
of the set of all jammed structures is a problem of great interest.
Besides the MRJ state, for example, one may wish to know
the lowest density jammed structure(s), which is an open
problem in two and three dimensions. We also note
that the concept of a jammed structure is particularly
relevant to the flow, or lack thereof, of granular media
\cite{Cat99,Lin00}.

In this paper, we will focus our attention on the question,
What does one really mean by a ``jammed'' hard-particle
system? The answer to this question is quite subtle and
a failure to appreciate  the nuances involved has
resulted in considerable ambiguity in the literature
on this question. Yet a precise definition for the
term ``jammed'' is a necessary first step before one can undertake
a search for jammed
structures in a meaningful way. We will show
that there is a multiplicity of definitions for jammed
structures. For simplicity and definiteness, we will restrict
ourselves to equi-sized $d$-dimensional hard spheres in
$d$-dimensional Euclidean space.
Of particular concern will be the cases
of equi-sized hard circular disks ($d=2$) and
equi-sized hard spheres ($d=3$).

\section{Definitions}

We will consider $N$ equi-sized $d$-dimensional hard spheres
of diameter $D$.
To begin, we will
assume that the $N$ particles are confined to a convex region
of $d$-dimensional Euclidean space of volume $V$ with impenetrable but possibly deformable
boundaries. The boundaries are assumed to be smooth on
the scale of the particle diameter.
Periodic boundary conditions will be mentioned
separately below. 

A particle in the system is individually
jammed if it cannot be translated
while holding fixed the positions of all of the other $N-1$ particles in 
the system. This means that a particle in the bulk
must have at least $d+1$ neighbor or wall contacts not
all of which are in the same ``hemisphere.''
A necessary condition for the entire system to be jammed is that 
each of the $N$ particles is individually jammed.
The system of disks shown in Fig. 1 meets this necessary condition, but all
are in contact with the impenetrable boundary, and consequently leave the 
interior of the system entirely vacant. Analogous three-dimensional examples
can also be identified, with all spheres jammed against the boundary, 
leaving the interior of the system totally unoccupied. 
While such unusual
cases may have some intrinsic interest, we leave them aside for present
purposes by requiring that at least one particle of the jammed configuration
not contact the boundary. This leaves open the possibility that the 
interior of some disk and sphere packings might display relatively large  
voids or cavities surrounded by a ``cage" of jammed particles.

Note that these minimal requirements mean that
there can be no ``rattlers''
(i.e., movable but caged particles) in the system. It  should be recognized that
jammed structures created in practice via computer algorithms \cite{Lu90}
or actual experiments may contain a small concentration of
such rattler particles, the precise concentration of which
is protocol-dependent.  
Nevertheless, it is
the overwhelming majority of spheres that compose the underlying ``jammed''
network that confers ``rigidity'' to the particle packing and, in any
case, the ``rattlers'' could be removed without disrupting
the jammed remainder.

We are now in a position to state our definitions of jammed
configurations.
A system of $N$  spheres is said to be a
\begin{enumerate}
\item {\it Locally jammed configuration} if the system
boundaries are nondeformable and each 
of the $N$ particles is individually jammed, i.e., it meets the 
aforementioned necessary condition.
\item {\it Collectively jammed configuration} if the system
boundaries are nondeformable and it is a locally jammed configuration
in which there can be no collective motion of
any contacting subset of particles that leads to unjamming.
\item {\it Strictly jammed configuration} if it is
collectively jammed and the configuration remains fixed
under infinitesimal virtual global
deformations of the boundaries. In other words, no global
boundary-shape change accompanied by collective particle
motions can exist, that respects the nonoverlap conditions.
\end{enumerate}
We emphasize that our definitions
do not exhaust the universe of possible distinctions, but
they appear to span the reasonable spectrum of possibilities.
It is clear that the second  definition is more restrictive
than the first and  the third
definition  is the most restrictive. It is crucial to observe
that the above classification scheme is dependent on
the type of boundary conditions imposed (e.g., impenetrable
or periodic boundary conditions) as well as the
shape of the boundary. 

\begin{figure}
\centerline{\psfig{file=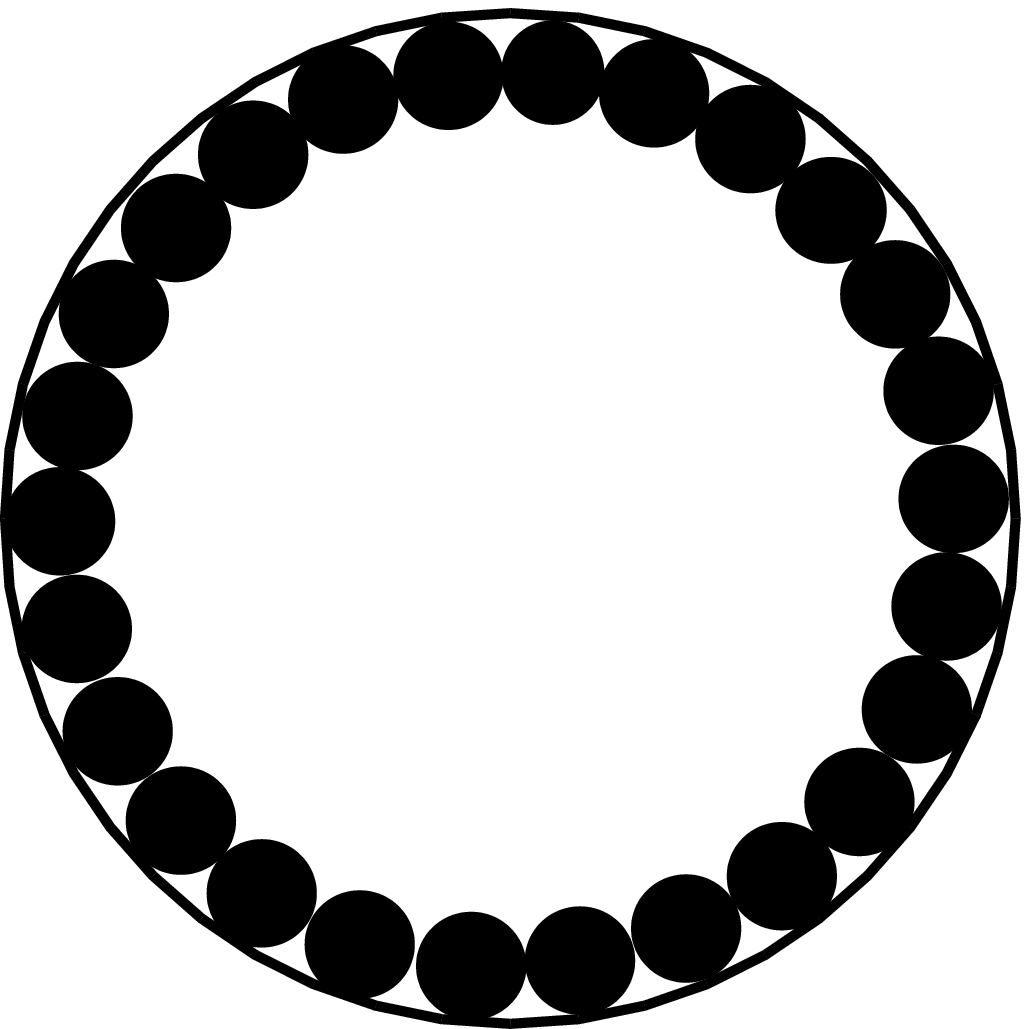,width=6.0in}}
\caption{A jammed system in which all of the particles
are in contact with the boundary.}
\label{circle}
\end{figure}

The collectively jammed definition was the one used
by Torquato et al. \cite{To00b} in their work
on the maximally random jammed state. Note that 
overall rotation of configurations in a circular
or spherical boundary can still leave the system
collectively jammed (see, for example, Fig. \ref{circle}).

Observe that the most restrictive definition of a jammed
structure, the {\it strictly jammed configuration},
is a purely kinematic one, i.e., we do not appeal
to a description of forces or stresses on the system.
However, one could choose to relate the concomitant
stresses on the boundaries to the deformations
via some appropriate constitutive relation.
For example, in the case that the stresses are linearly related
to the strains,
Hooke's law for linear elasticity would apply and the system would
be characterized generally by 21 elastic
moduli in three dimensions. In the case of
an elastically isotropic packing, two
elastic moduli would characterize the system:
the bulk modulus $K$, relating isotropic compressive
stresses to corresponding volumetric strains (deformations), and
the shear modulus $G$, relating shear stresses
to corresponding volume-preserving strains.
Thus, in this latter instance, a strictly jammed configuration
is characterized by infinite bulk and shear moduli.

An interesting characteristic of a jammed packing of spheres
that has not been studied to our knowledge is its
``{\it rigidity percolation threshold}.'' Rigidity percolation
has been studied on lattice networks to understand the mechanical 
properties of network glasses, 
for example \cite{Thorpe85,Mo99,Avr00}. Consider a triangular
net of mass points connected by nearest-neighbor central
forces. The system is stable and elastically isotropic,
and therefore is characterized by the elastic moduli $K$
and $G$. If bonds are randomly removed with probability $1-p$,
then both $K$ and $G$ vanish at some critical value 
$p^*$ between 0 and 1 called the rigidity percolation threshold.

The notion of rigidity percolation can be extended to describe jammed sphere packings.
Importantly, in doing so, we do not have to appeal to notions of elasticity.
Consider a jammed system that meets one of the three
aforementioned definitions. Begin a process whereby 
spheres are sequentially removed by some selection
process with a random element. The rigidity
percolation threshold is the sphere volume fraction $\phi^*$ at which
the system ceases to be jammed according to one of our three definitions.
Thus, the value of $\phi^*$ will generally vary for a given
structure depending on what is meant by a jammed configuration.
For example, for an initial lattice at packing fraction $\phi$, 
the value of $\phi^*$ will increase as the jamming criterion
changes from the least restrictive
(locally jammed) to the most restrictive (strictly jammed). In general,
$\phi^*$ must lie in the interval $(0,\phi]$. 
We believe that this generalization of rigidity percolation
will be especially useful in characterizing random jammed
packings.

 Boundary conditions play an essential role in packing problems. 
 Although the principal focus of the present exposition concerns the case 
 of impenetrable boundaries (on account of their physical significance), 
 we recognize that periodic boundary conditions are often applied in a 
 wide range of many-particle theories and numerical simulations.  Note 
 that in the present context periodic boundary conditions are 
 substantially less confining than are impenetrable-wall boundary 
 conditions, when either could be applied to a given finite particle 
 packing.  Hence, the classification scheme defined above may have an 
 outcome that hinges sensitively upon which of these alternatives 
 applies.  The simple cubic lattice offers an example; it is collectively 
 jammed with rigid walls, but only locally jammed with periodic boundary 
 conditions.

\section{Classification of Some Ordered Lattice Packings}

\begin{figure}
\centerline{\psfig{file=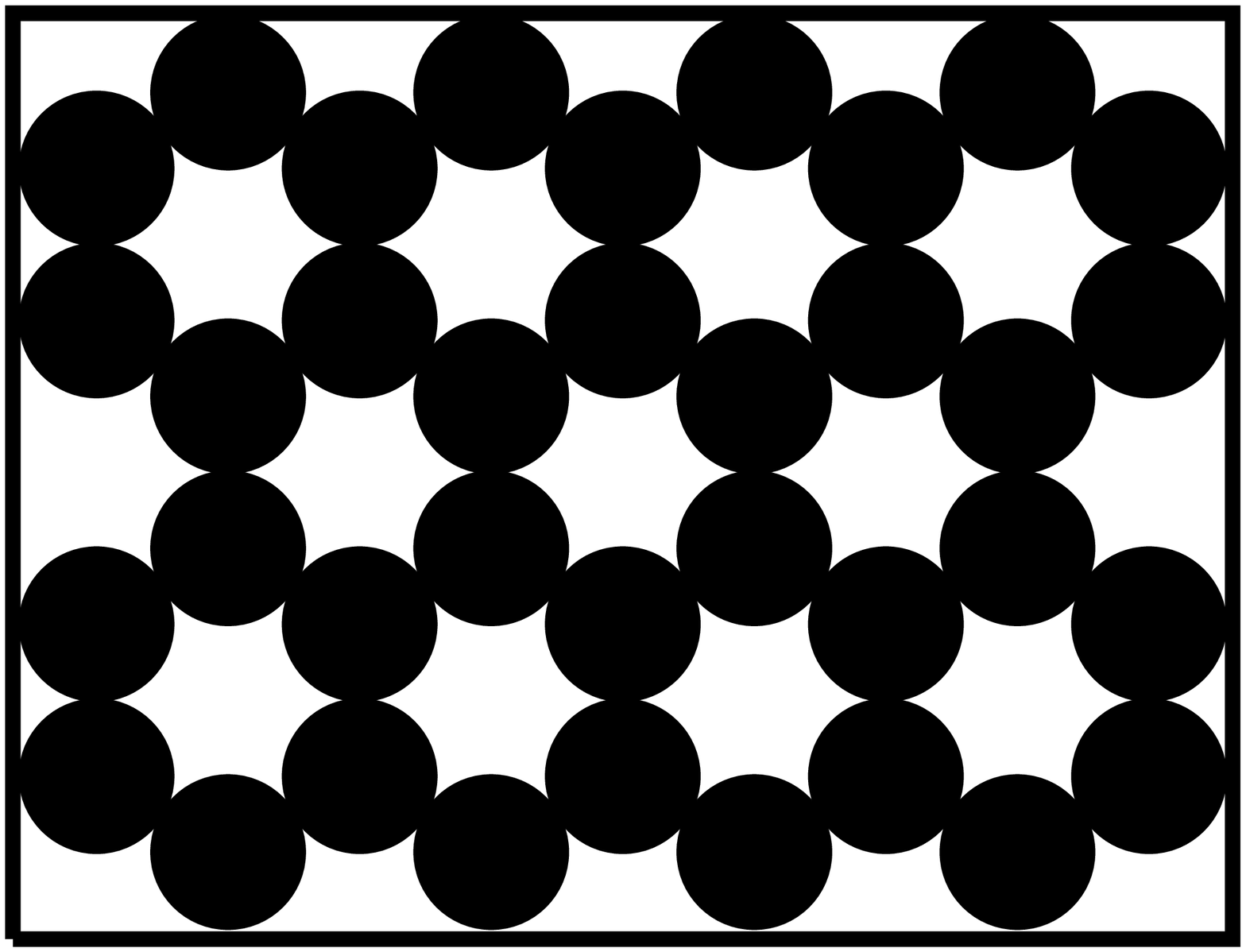,width=3.2in}\hspace{0.2in}
\psfig{file=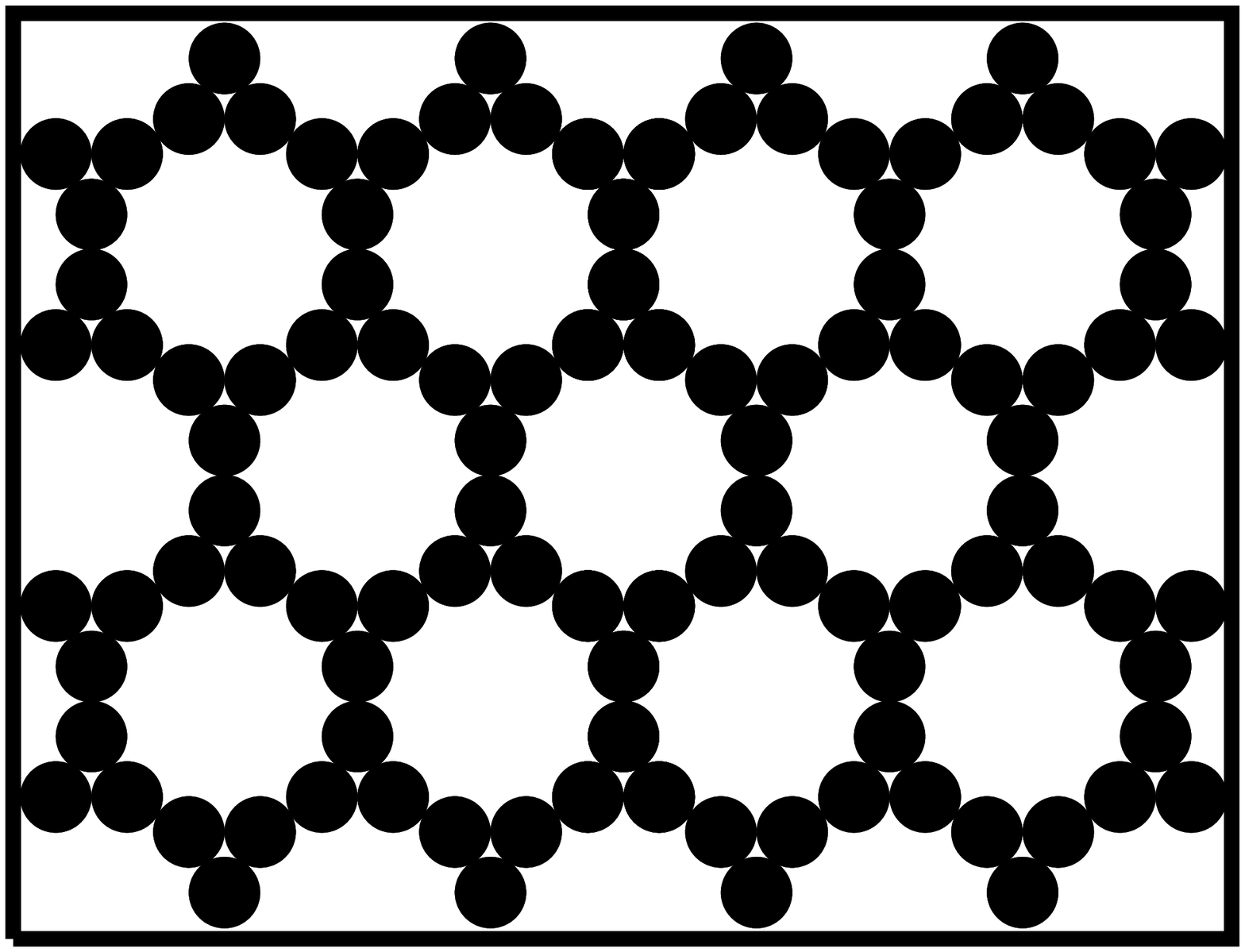,width=3.2in}}
\caption{Left panel: The honeycomb (hexagonal) lattice can be
made locally jammed but not collectively jammed with a hard
rectangular boundary. Right panel: The packing that
results by placing within each disk in the left panel three
smaller disks that are locally jammed and then removing the
larger disks. This procedure is a means of creating low-density
jammed structures in the infinite-volume limit.}
\label{honey}
\end{figure}

In this section we categorize some common two-
and three-dimensional lattice packings according 
to whether they are locally jammed, collectively
jammed, or strictly jammed. In all cases we assume 
impenetrable but possibly deformable system
boundaries. We begin with
two-dimensional lattices within 
commensurate rectangular boundaries. The three-fold coordination
of a honeycomb (hexagonal) lattice is sufficient to make
this structure locally jammed (see Fig. \ref{honey}a).
Note that the graph that results by drawing lines
between nearest-neighbor centers is a hexagonal
tiling of the plane for an infinite system. 
In the infinite-volume limit, the packing fraction $\phi=\pi/(3\sqrt{3})\approx 0.605$.
However,  the honeycomb lattice is  not
collectively jammed, since an appropriate collective rotation
of six particles that are situated on the
sites of any of the hexagons in the bulk will destabilize
the structure. Thus, the honeycomb lattice
is also not strictly jammed.
Note that by an appropriate placement of three circular
disks of diameter $\sqrt{3}D/4$ in each of the
original disks of diameter $D$ in Fig. \ref{honey}a, the packing that results
upon removal of the larger disks is locally jammed
at the packing fraction $\phi=3\pi/(4\sqrt{3})\approx 0.340$ in the infinite-volume 
limit (see Fig. \ref{honey}b).   

\begin{figure}
\centerline{\psfig{file=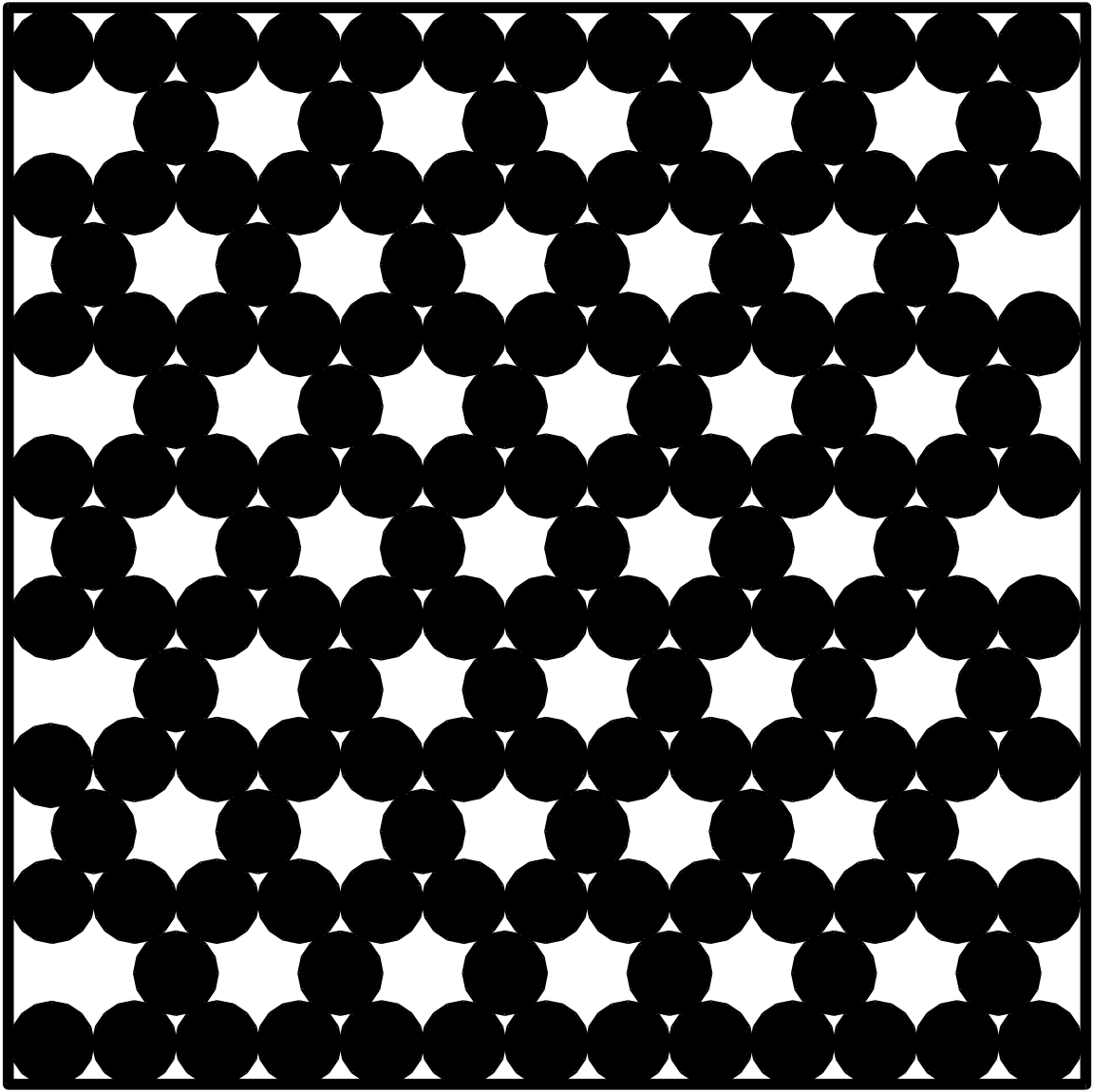,width=6.0in}}
\caption{The Kagom{\'e} lattice is not locally jammed with a hard
rectangular boundary. However, when properly situated within a container
with a regular triangular- or hexagonal-shaped boundary, it
can be made to be strictly jammed.}
\label{kagome}
\end{figure}

\begin{figure}
\centerline{\psfig{file=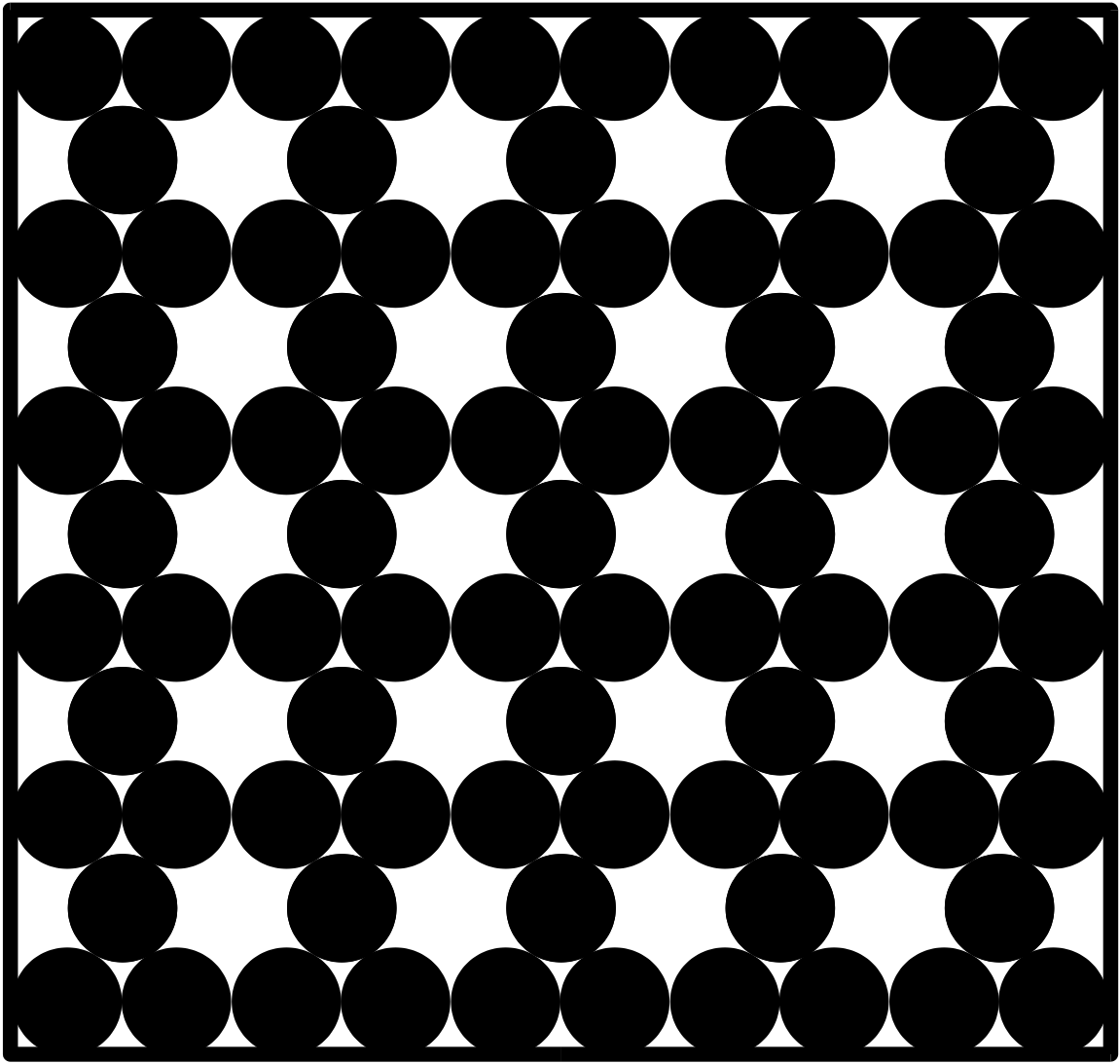,width=6.0in}}
\caption{An example of a two-dimensional packing that is collectively
jammed with a hard rectangular boundary.}
\label{frank}
\end{figure}

The square lattice 
is both locally and collectively jammed, but
it is not strictly jammed, since a shear
deformation (not an isotropic deformation)
will destabilize the packing. The Kagom{\'e} lattice
is not locally jammed in a rectangular container
because certain particles along
the vertical walls may be moved, leading
to an instability (see Fig.~\ref{kagome}). However,
the Kagom{\'e} lattice becomes strictly jammed
if it is appropriately situated within
a container with either regular triangular- or
hexagonal-shaped boundaries. Note that for an infinitely large system, it
has a packing fraction of $\phi=3\pi/(8\sqrt{3})\approx 0.680$.
The triangular lattice is strictly jammed. It
has a packing fraction of $\phi=\pi/\sqrt{12}\approx 0.907$.
The classification of all of the aforementioned lattices 
is summarized in Table~\ref{class}. An example of a two-dimensional 
collectively jammed
lattice with an appreciably lower packing fraction 
than the triangular lattice 
is shown in Fig. \ref{frank}. This four-coordinated
lattice, built from the triangular lattice with
one-fourth of the disks missing, was considered
by Lubachevsky, Stillinger, and Pinson \cite{Lub91}.

Now let us consider some three-dimensional lattices
within a cubical container.
The tetrahedrally coordinated diamond lattice is 
the three-dimensional analogue of the
honeycomb lattice. It is locally jammed,
but is not collectively jammed, and therefore
not strictly jammed. The simple cubic lattice
is the three-dimensional analogue of the square
lattice; it is both locally and collectively jammed, but
it is not strictly jammed, since a shear
deformation (not an isotropic deformation) in the $(100)$
planes will destabilize the packing. The same characterization
is true for the  body-centered
cubic (BCC) lattice as for the simple-cubic lattice.
The BCC lattice is not strictly jammed because
a shear deformation in the $(110)$ planes
will destabilize the packing.
The face-centered cubic lattice is strictly jammed.
The hexagonal close-packed lattice is  strictly jammed
if it the container boundary is a hexagonal prism but
it is only locally jammed for a cubical boundary.

\section{Discussion and Future Work}

Let us now turn our attention to the practical determination
of our three different definitions for jammed configurations.
It is clear that the criteria for a locally jammed configuration
will be the easiest to implement in a computer simulation.
In two dimensions, one must determine whether each
disk is locally jammed, i.e., whether each disk
has at least three contacting neighbors that do not
all lie in a semicircle 
surrounding the particle of concern.
In three dimensions, one must ascertain whether each sphere
has at least four contacting neighbors that do not
all lie in a hemisphere surrounding the particle.
Error enters the search algorithm because one must choose an acceptable tolerance 
for the nearest-neighbor distance to determine
whether a neighbor is indeed in contact with
the reference particle.

The determination of whether the system is collectively
jammed is considerably more difficult, especially for a random
system. An approximate means to test that the  system is collectively
jammed is to shrink the particle sizes uniformly by a very small 
amount, give the particles some random initial velocities,
and follow the system dynamics using a molecular dynamics
simulation technique. If the particle configuration
effectively does not change after a sufficiently long period of time,
the system can be regarded to be collectively
jammed. This procedure has been used
by Lubachevsky, Stillinger, and Pinson \cite{Lub91} to
ascertain whether their particle systems were ``stable.''
Such a stochastic approach is intended to discover if the particle 
configuration considered contains polygons of contacting neighbors whose 
simultaneous displacements initiate local unjamming.  A desirable 
objective for future research is the design of a more efficient 
discovery procedure for these multiparticle unjamming motions.

Interestingly, the determination of whether a particle packing
is strictly jammed may be relatively straightforward given
that it is collectively jammed. The basic idea is to transform
the collectively jammed particle packing into an equivalent Delaunay graph or network.
Roughly speaking, the Delaunay network is the polyhedral graph that results by
drawing lines between nearest-neighbor centers in the packing \cite{Au91}.
Once this equivalent network is determined, one can exploit
well-developed engineering techniques to analyze 
the stability of truss-like structures \cite{Pel93}. Specifically,
overall tractions  are imposed on the boundary
of the network and the stability analysis is reduced 
to a  well-defined linear algebra problem. If the network does not deform under these
boundary conditions, then it is stable, or, equivalently,
the packing is strictly jammed. If the network deforms,
then the packing is not  strictly jammed.
To our knowledge, the use of such techniques to characterize
jammed packings would be new.

The classification of  random packings of $d$-dimensional spheres
according to our criteria defined above possesses direct relevance to
the ongoing search for  maximally random jammed (MRJ) states.
  Even for a
given choice of scalar order parameter $\psi$, the maximally random
jammed state can be expected to depend nontrivially on which of the
three jamming definitions (locally, collectively, or strictly jammed)
has been imposed.  It seems likely that there is a wide class of random
packings that satisfy both the locally jammed and collectively jammed
criteria.  However, it is not clear whether random packings can be
strictly jammed with more than vanishingly small probability.  In other
words, it may be very unlikely to find collectively jammed
configurations that are able to resist all shear deformations.  As noted
earlier, our suggested generalization of the rigidity percolation
concept may prove valuable in identifying strictly jammed structures,
and in characterizing those local geometric attributes which allow them
to resist shear.

\section{Concluding Remarks}

We have shown
that there is a multiplicity of generation, selection, and
classification procedures for jammed
configurations of identical $d$-dimensional hard spheres.
In particular, we have given three different definitions 
for jammed configurations: (1) locally jammed configuration; (2) collectively jammed
configuration; and (3) strictly
jammed configuration. Importantly, the particular classification
of a random packing depends crucially
on the type of boundary conditions imposed as well
as the shape of the boundary.
We also have shown 
how the concept of rigidity percolation, previously applied
to understand the mechanical properties
of network glasses, can be generalized to 
characterize hard-sphere packings even further. 
We have categorized common ordered lattices according
to our definitions and discuss implications for
random disk and sphere packings. Thus, we see that the characterization
of jammed hard-particle packings is inherently nonunique
and that the choice one makes is ultimately problem-dependent. Finally, we have
discussed the practical implementation of our three different definitions 
for jammed configurations.

\newpage
\bigskip
\bigskip
{\bf Acknowledgments}

The authors are pleased to acknowledge  the stimulating influence
of discussions with, and papers by, Howard Reiss concerning 
hard-particle statistical phenomena.
The authors are grateful to Juan Eroles for creating the figures
in this paper. 
S. T. was supported by the 
Engineering Research Program of the Office of Basic Energy Sciences at
the Department of Energy and the Petroleum Research Fund as administered
by the American Chemical Society.

\newpage
\normalsize

\newpage

\newpage

\newpage

\newpage

\begin{threeparttable}
\caption{Classification of 
some of the common jammed ordered lattices of equi-sized spheres
in two and three dimensions, where
$Z$ denotes the coordination number and $\phi$ is
the packing fraction for the infinite lattice.
Here hard boundaries are applicable: in two dimensions
we use commensurate rectangular boundaries
and in three dimensions we use a cubical
boundary, with the exception of the hexagonal close-packed 
lattice in which the natural choice is a hexagonal prism.}
\small
\begin{tabular}{c|c|c|c} \hline \hline 
Lattice  &Locally jammed & Collectively jammed &  Strictly
jammed \\ \hline \hline
Honeycomb ($Z=3$, $\phi \approx 0.605$) &yes & no&no\\ \hline
Kagom{\'e} ($Z=4$, $\phi \approx 0.680$)&no\tnote{a} &no\tnote{a} & no\tnote{a}\\ \hline
Square ($Z=4$, $\phi \approx 0.785$)&yes &yes & no\\ \hline
Triangular ($Z=6$, $\phi \approx 0.907$) &yes& yes&yes\\ \hline \hline
Diamond ($Z=4$, $\phi \approx 0.340$)&yes&no&no\\ \hline
Simple cubic ($Z=6$, $\phi \approx 0.524 $)&yes& yes&no \\ \hline
Body-centered cubic ($Z=8$, $\phi \approx 0.680$)&yes&yes&no\\ \hline
Face-centered cubic ($Z=12$, $\phi \approx  0.741 $) &yes&yes&yes\\ \hline
Hexagonal close-packed ($Z=12$, $\phi \approx 0.741$) &yes&yes&yes\\ \hline \hline
\end{tabular}
\begin{tablenotes}
\item[a] With appropriately placed regular triangular- or
hexagonal-shaped boundaries, the Kagom{\'e} lattice
is locally, collectively and strictly jammed.
\end{tablenotes}
\label{class}
\end{threeparttable}

\end{document}